\documentstyle[12pt,bezier]{article}
\newcommand{\beq}{\begin{equation}}
\newcommand{\beqa}{\begin{eqnarray}}
\newcommand{\eeq}{\end{equation}}
\newcommand{\eeqa}{\end{eqnarray}}

\def\eps{\epsilon}
\begin{document}
\begin{titlepage}
\pagestyle{empty}
\baselineskip=21pt
\rightline{UMN-TH-1259-94}
\rightline{TPI-MINN-94/20-T}
\rightline{hep-ph/9405365}
\rightline{revised August 1994}
\vskip .2in
\begin{center}
{\large{\bf Quark Diffusion and Baryogenesis\\
at the Electroweak Phase Transition}}
\end{center}
\vskip .1in
\begin{center}
James M.~Cline\\
{\it School of Physics and Astronomy, University of Minnesota}

{\it Minneapolis, MN 55455, USA}

\vskip .2in

\end{center}
\vskip 1in
\centerline{ {\bf Abstract} }
\baselineskip=18pt
\noindent
Baryogenesis at the electroweak phase transition may take place through
CP-violating reflections of quarks from expanding bubbles of the broken
symmetry phase. We formulate and approximately solve the transport equations
for the reflected quark asymmetries.  The results are comparable to a previous
Monte Carlo calculation and allow for analytic estimates of the baryon
asymmetry $\Delta B$ when the bubble walls have a small velocity. Our method
predicts no velocity-dependence for $\Delta B$ in the small $v$ limit, unlike
results obtained from a simplified treatment based on the diffusion equation.
\end{titlepage}
\baselineskip=18pt

There is currently much interest in the possibility that the baryon asymmetry
of the universe was created during the electroweak phase transition, using its
first-order nature as the means for departure from thermal equilibrium, as is
required in any theory of baryogenesis [1-4]. There are two paradigms,
depending on whether the bubbles of true vacuum which nucleate in the symmetric
phase during the transition have relatively thin or thick walls.  In the
thin-wall regime which is investigated here, CP-violation in the Higgs sector
causes an asymmetry between fermions and antifermions reflected from the bubble
walls. To leading order in $\alpha_{\rm weak}$ it consists of equal and
opposite excesses of right-handed and left-handed fermions, due to CPT
conservation, so that the net baryon or lepton number in the disturbance is
zero, but its chirality is nonzero.  This asymmetry is subsequently converted
to a baryon asymmetry by sphaleron interactions.  The expanding wall overtakes
the baryon asymmetry and incorporates it into the true vacuum phase where it is
preserved because of the freezeout of sphalerons.

To compute the resulting baryon asymmetry in this scenario, it is necessary to
know the distribution functions of the fermions and antifermions in front of
the wall in some detail.  So far several treatments have been given, including
a Monte Carlo simulation of the diffusion of reflected quarks \cite{CKN}
(hereafter called CKN), and the diffusion equation \cite{JPT}. Our aim is to
complement these approaches by starting with the exact Boltzmann equation and
solving for the distributions with the help of a few reasonable approximations.
We will find the profile for the reflected quark asymmetry, estimate the effect
of nonvanishing domain wall thickness on the baryon asymmetry, and show that
the present method disagrees with the diffusion equation approach concerning
the velocity dependence of the final result.

Our starting point to compute the transport of reflected quarks with
distribution functions $f_i$ is the Boltzmann equation,
\beq
        \left[\partial_t + {{\bf p}\over E}\cdot \partial_{\bf x}
        \right]f_i({\bf x, p}) = C_i
\label{boltzmann}
\eeq
where $C_i$ is the collision term.\footnote{We have ignored force terms that
would give rise to the screening of hypercharge \cite{screening} because any
deviations $\delta\!f_i$ due to screening would be of the form $y_i \phi$,
where $y_i$ is the hypercharge of particle $i$ and $\phi$ is the hypercharge
potential to be screened.  Such a deviation would have no effect on the biasing
of sphaleron interactions because they conserve hypercharge.} The latter is the
standard difference between two terms, which scatter the particle of interest
respectively into and out of an element of phase space,
\beqa
        C_i &=& \sum_{\rm channels}
        {1\over 2E_i}\int d\Pi_1 d\Pi_2 d\Pi_3\Bigl\{
        (2\pi)^4 \delta^4(p_1+p_2-p_3-p)|{\cal M}_{12\to 3p}|^2
        f({\bf p_2}) f_i({\bf x,  p_1}) \nonumber\\
        && \qquad\qquad\qquad\qquad- ({\bf p_1\leftrightarrow \bf p})
	\Bigr\};\quad d\Pi_j\equiv d^3 p_j/(16\pi^2 E_j),
\label{collision}
\eeqa
ignoring Pauli blocking factors.  $f(\bf p)$ is the
nearly-equilibrium distribution function for the quark or gluon off of which
the species $i$ of interest is scattering.  Of course there is one of
approximately 20 channels where the former is also the quark of flavor $i$, in
which case $f_i$ would appear quadratically, but ignoring this makes only a
small error.  We have also omitted decay processes because the thermal masses
of the gauge bosons are below the threshold for producing two thermal quarks,
and the rate for Higgs boson decays into fermions is smaller than scatterings
mediated by gauge bosons, due to the large number of scattering channels.

Because the scattering processes are dominated by low momentum transfer, due to
the near masslessness of exchanged gauge bosons in the thermal plasma, it is
possible to simplify the collision term using the Fokker-Planck approximation
\cite{Landau}.  The key observation is that in the absence of a thermal mass
$m_g$ for the gluon, ${C}_i$ would diverge logarithmically in the infrared
because the matrix element has the leading behavior $|{\cal M}|^2 \sim g_s^4
s^2/(t-m^2_g)$ in terms of Mandelstam variables.  We will isolate the logarithm
and discard the corrections that would vanish as $m_g\to 0$. Let $\bf p$
denote the three-momentum of the scattered quark, and define $\bf p_2 = \bf
p\,'$; $\bf p_3 = \bf p\,' + \bf q$; so $\bf p\,'$ is the momentum of the
background quark or gluon and $\bf q$ is the momentum transfer.  By
substituting $\bf p_1 = \bf p+\bf q$ in the first term on the right hand side
of (\ref{collision}) and $\bf p_1 = \bf p-\bf q$ in the second term, and doing
the integral over $d\Pi_3$ using the momentum-conserving delta function, we get
\beqa
\label{difference}
   {C}_i &=& {1\over 16(2\pi)^5}\int d^3 q \left\{ G({\bf p}+{\bf q},
	{\bf q)} - G({\bf p},{\bf q}) \right\}; \\
	G({\bf p},{\bf q}) &=& {f_i({\bf x}, {\bf p})\over |{\bf p}||{\bf p}
	-{\bf q}|}\int{d^3p'f(p')\over |{\bf p\,'}||{\bf p\,}'+{\bf q'}|}
	\delta(|{\bf p}|+|{\bf p\,'}|
	-|{\bf p}-{\bf q}| - |{\bf p\,'}+{\bf q}|)
	|{\cal M}_{p,p'\to p-q,p'+q}|^2\nonumber
\label{Gpq}
\eeqa
as the contribution from a single scattering channel; we will sum over channels
in the end. The next step is to approximate the integrand of (\ref{difference})
as
\beq
	\left[{\bf q}\cdot \partial_{\bf p} +\frac12
	({\bf q}\cdot\partial_{\bf p})^2 \right] G({\bf p, q})
\label{taylor}
\eeq
by Taylor-expanding.  This is a good approximation since we expect the
scattering to be dominated by small momentum transfer.  One finds that only
these first two terms are logarithmically sensitive to the gluon mass; the
higher ones are convergent.  It is farily straightforward to evaluate
them, with the result that the collision term can finally be expressed as
\beq
        C_i = D \partial_{\bf p}\cdot \left(\partial_{\bf p}f_i({\bf x, p})
        + {\beta{\bf p}\over E}f_i({\bf x, p}) \right).
\label{FP}
\eeq
This form is quite general and appears in nonrelativistic scattering processes
as well \cite{Landau}.

The diffusion coefficient emerging from this reduction is
\beq
        D = {20 \zeta(3)\over \pi}\alpha_s^2 T^3 \int_{m^2_g} {dq^2\over q^2},
\label{diffusion}
\eeq
where the factor of $20$ comes from weighting the multiplicity of different
scattering channels by their respective squared matrix elements and thermal
statistical factors ($8\times 1\times 1$  for gluons, $18\times 4/9\times 3/4$
for quarks and the same for antiquarks).  To evaluate the logarithm we take the
Debye screening mass $m^2_g = 8\pi\alpha_s T^2$ for the lower limit, the
thermally averaged, squared, center of mass energy $s\cong 20 T^2$ for the
upper limit, and $\alpha_s = 0.1$, with the result $D = 0.16 T^3$.  Since we
have ignored weak interactions, $D$ is the same for both quark chiralities.
Note that $D$ differs in dimension and meaning from the usual diffusion
coefficient, since it multiplies derivatives with respect to momentum rather
than position.

It is convenient to work in the rest frame of the wall, where the distribution
will be stationary so that the time derivative can be neglected. In this frame
we should Lorentz-transform the momentum variables in (\ref{boltzmann}) and
(\ref{FP}), but this complicates the solution of the transport equation, so we
will take the limit of small wall velocities $v$, where the difference
between the momenta in the rest frame of the universe and that of the bubble
wall can be neglected.  It will turn out that the asymmetry we seek between
quarks and antiquarks is of order $v$ so that simply dropping the time
derivative in eq.~(\ref{boltzmann}) amounts to ignoring higher order
corrections in $v$.

The term linear in derivatives in eq.~(\ref{FP}) is cancelled if we write $f_i
= e^{-\beta E/2} \hat f_i$.  Also, since we are interested in the asymmetries
between particles and antiparticles, $\delta\!f \equiv f_i - f_{\bar\imath}$,
we must subtract from eq.~(\ref{boltzmann}) the same equation for
$f_{\bar\imath}$.  The solution will have the form
\beq
        \delta\!f(p,z) = e^{-\beta E/2}\sum_{n=0}^\infty\int_{k_0}^\infty
	dke^{-k|z|}
        \left( \theta(z) A^+_{kn} g^+_{kn}(p)
        + \theta(-z) A^-_{kn} g^-_{kn}(p)\right).
\label{soln1}
\eeq
The differential equation for $g_{kn}^\pm$ is
\beq
        \left(\partial_p^2 -{\beta^2 p^2\over 4 E^2} + {\beta\over 2}
        \left({3\over E} - {p^2\over E^3}\right)
        \mp {k p_z\over DE}\right)g_{kn}^\pm = 0,
\label{FP2}
\eeq
where one must remember that $E = p$ for $g_{kn}^+$ but $E=(p^2+m^2)^{1/2}$ for
$g_{kn}^-$ because the electroweak symmetry is broken only to the left of the
wall.  However, we will solve (\ref{FP2}) only to leading order in the quark
mass $m$.  Then, similarly to our neglect of the wall velocity in the
Fokker-Planck equation, we can set $E = p$ to discard higher order corrections
in $m$, since the solution $\delta\!f$ is homogeneous in a source term that
vanishes as $m\to 0$. Throughout this paper $m$ is understood to be the quark
mass in the broken symmetry phase at the critical temperature $T_c$, which is
smaller than the usual mass because the Higgs field VEV is smaller at $T_c$
than at $T=0$.

Using parabolic coordinates $p_z = \frac12(\xi-\eta)$, $p = \frac12(\xi+\eta)$,
the massless limit of eq.~(\ref{FP2}) is separable and it can be solved
exactly:
\beqa
\label{soln2}
        g^+_{kn}(\xi,\eta) &=& e^{-a_+\eta/2+ia_-\xi/2}\ _1F_1(-n;1;a_+\eta)
        \ _1F_1(c_{kn};1;-ia_-\xi); \\
        a_\pm &=& (k/D \pm \beta^2/4)^{1/2};\quad
        c_{kn} = 1/2 - (i/a_-)\left(\beta/2-(n+
        {\scriptstyle{ 1\over 2}})a_+
        \right),\nonumber
\eeqa
and $g^-_{kn}(\xi,\eta) = g^+_{kn}(\eta,\xi)$.  The requirement that $n$ be an
integer in eq.~(\ref{soln2}) is to insure good behavior as $\eta\to\infty$.
The functions $g_{kn}^\pm$ form a complete set for $k\geq k_0\equiv \beta^2
D/4\cong 0.04T$.   Note that the spatial extent of the diffusion layer will
therefore be given by the distance scale $k_0^{-1}$, from eq.~(\ref{soln1}).

Let us now discuss the boundary conditions for the distributions, which express
the conservation of flux at the bubble wall.  For example, since left-handed
particles are reflected into right-handed particles and vice versa (because of
angular momentum conservation), the distribution function for left-handed
quarks satisfies
\beq
        f_{t_L}^+(p_z) = {\cal R}_{R\to L} f_{t_R}^+(-p_z)
        + {\cal T}_L f_{t_L}^-(p_z)\qquad (p_z>m),
\eeq
where the superscripts indicate that $f_i$ is evaluated at $z=0+$ or $z=0-$
with respect to the wall, ${\cal R}_{R\to L}$ is the probability for the
reflection of right- to left-handed quarks, and ${\cal T}_L$ is the
transmission probability for left-handed quarks.  Subtracting the distribution
function of the antiparticle to obtain $\delta\!f^+\equiv f_{t_L}^+ -f_{\bar
t_L}^+$, remembering that $t_R$ has the corresponding asymmetry $-\delta\!f^+$,
and doing the same for the distributions on the $z<0$ side of the wall, one
obtains
\beq
        \delta\!f^+(\pm p_z)-\delta\!f^-(\pm p_z) = \mp{\cal R}_0
        \left(\delta\!f^+(-p_z) +\delta\!f^-(p_z)\right) + S(p_z)
        \qquad (p_z > m),
\label{bc1}
\eeq
where ${\cal R}_0$ is the mean value of the two reflection coefficients
${\cal R}_{R\to L}$, ${\cal R}_{L\to R}$, and the source term $S(p_z)$ depends
on their difference $\Delta{\cal R}$, and on that of the unperturbed
distribution functions:
\beqa
        S(p_z) &=&  \left(f^-(p_z) - f^+(-p_z)\right)\Delta{\cal R}(p_z)
		\nonumber\\
               &\cong& 2\beta v p_z e^{-\beta p}\Delta{\cal R}(p_z).
\label{source}
\eeqa
Here we used the equilibrium distribution functions $f(\pm p_z) \cong
e^{-\beta(p \pm vp_z)}$ on either side of the wall, expanding to first order in
the velocity $v$.  The difference between positive and negative $p_z$ is due to
our working in the rest frame of the bubble wall.

As one would intuitively expect, the solution to the boundary conditions
(\ref{bc1}) requires that $\delta\!f^-(p_z) = - \delta\!f^+(-p_z)$: there is a
dipole layer of left-handed quark excess at the bubble wall, since the net
number of chiral quarks is conserved and remains zero during the process of
reflection.  This means that $A^-_{kn}= -A^+_{kn}$, and the boundary condition
becomes an integral equation for $A^+_{kn}$, which can be solved using the
orthogonality of the eigenfunctions $g_{kn}^\pm(\xi,\eta)$:
\beq
        A^+_{kn} = N^{-1}_{kn}\int_0^\infty d\xi \int_0^\xi d\eta (\xi - \eta)
        e^{\beta p/2} S(p(\xi,\eta)) \left[g^+_{kn}(\xi,\eta) -
        g^+_{kn}(\eta,\xi) \right].
\label{Akn}
\eeq
Here $N_{kn}$ is the normalization of the orthogonality integral of the
$g^+_{kn}$'s, $N_{kn} = (D/8a_+) \sin(\pi c_{kn})$.

We thus have an approximate solution for the left-handed quark asymmetry in the
symmetric phase, which enables us to compute the baryon asymmetry that
sphalerons will be driven to produce.  A convenient measure is the
baryon-to-entropy ratio. In {CKN} it is found that in a theory with $n$
Higgs doublets (we will assume $n=2$),
\beq
   {\rho_B\over s} \cong {3\times 10^{-7}\kappa\over (2n+1) v} \beta^2\delta n,
\label{BAU}
\eeq
where $\kappa$ is the dimensionless constant ($\kappa\le 1$) characterizing the
rate of sphaleron interactions in the symmetric phase, and $\delta n$ is the
number of reflected quarks per unit area in front of the wall,
\beq
        \delta n = \int d^3 p \int_{z_0}^\infty dz\, \delta\! f(p,z).
\label{dn0}
\eeq
Notice that in our approximation of small wall velocities, $\delta n\sim S(p)
\sim v$ so the baryon asymmetry is independent of $v$ in this limit.

The lower limit $z_0$ of the spatial integration in eq.~(\ref{dn0}) requires
some explanation since it does not appear in CKN: due to the finite thickness
of the domain wall, the spatial region in which baryogenesis takes place is
reduced.  It was estimated in CKN that the baryon number violating interactions
effectively turn off, in going from the symmetric to the broken phase, when the
Higgs field reaches a value of $g/4\pi$ of its full VEV.  For a profile of the
form $\phi_c(1+\tanh(z/\Delta))$ this occurs at $z=z_0\cong 1.5\Delta$.  If the
Higgs potential is parametrized as $\lambda \phi^2(\phi-\phi_c)^2$ at the
critical temperature, then $\Delta = (2/\lambda)^{1/2} \phi_c^{-1}$, and the
experimental lower limit on the mass of the Higgs boson gives $\sqrt{\lambda} >
0.085$.   We assume that this bound is saturated, as well as the bound $\phi_c
> T_c$, which assures that sphaleron interactions in the broken phase are
sufficiently suppressed to avoid washing out the baryon asymmetry on
cosmlogical timescales.   It is convenient to express $z_0$ in units of the
characteristic width of the diffusion layer.  The result is $z_0 k_0 = 1.0$.

In order to evaluate the density $\delta n$ of reflected quarks in front of the
wall, one must assume a form for the reflection asymmetry $\Delta R(p_z)$.  By
analyzing existing results \cite{CKN}\cite{Fun} it can be shown that $\Delta
R$ is rather well fitted by the form
\beq
	\Delta R(p_z) = N\theta(\eps)\eps^\alpha e^{-\eps/w};
	\quad \eps\equiv p_z - m.
 \eeq
The parameters $N$, $w$ and $\alpha$ depend only on the quark mass,
and $\alpha$ is neligible for a particle as heavy as the top quark \cite{CKV}.
Then, although the explicit expression for $\delta n$ is rather complicated, it
takes a simple expression in the limit that the temperature is much larger than
any other mass scales in the problem,
\beq
	\delta n = {N w v\over D\beta}
	\left\{\begin{array}{cc} 31.7 w^3, & T\gg w \gg m_q\\
				 14.3 m_q^3, & T\gg m \gg w \end{array}\right\}
				 s(z_0),
\label{dn}
\eeq
after numerically evaluating the sum/integral appearing in eq.~(\ref{soln1}).
The suppression factor $s(z_0)$ due to finite wall thickness is shown in
figure 1.  For the value of $z_0$ determined above we see that $s(z_0)$ is
approximately 0.12, which is a source of suppression that has not previously
been taken into account.

To compare to the Monte Carlo computation of $\delta n$ by {CKN}, we can use
their determination of the reflection asymmetry in our result, eq.~(\ref{dn}).
For example, in the most favorable case when the width of the wall is assumed
to be $m^{-1}$ and CP violation is maximal, they find a reflection asymmetry
with $N = 0.68$ and $w = 0.44 m$, from which we obtain  $\rho_B/s = 1.9\times
10^{-7}\kappa(m/T_c)^4$.  Our result is quite close to that of CKN for small
wall velocities ($v = 0.1$) and $m/T_c < 1$, but for larger $m$ it begins to
exceed their result.  Of course in this region our approximations are no longer
valid, but for a top quark of mass 170 GeV the Yukawa coupling is $y\cong 1$,
and $m/T_c = y\phi_c/T_c \cong 1$.

We have computed the shape of the profile $\delta n(z)$ in the same
approximation as used to obtain the integrated value of $\delta n$ above. It is
shown in figure 2, where one sees that at a distance of $k_0^{-1}= 25/T$ it has
dropped to 5\% of its maximum value, and at $2k_0^{-1}$ to $0.8$\%.  Near the
wall it falls off faster than an exponential, but asymptotically approaches
exponential decay far from the wall. It is interesting to compare the width of
the profile with the mean free path of quarks due to QCD scatterings, which we
have computed to be $\lambda = 3 T^{-1}$.  Figure 1 indicates that the profile
falls by a factor of $e$ in a similar distance, $4 T^{-1}$, comparable to the
diffusion length obtained using the position-space formulation of the diffusion
problem \cite{Heis}.

The behavior of $\delta n(z)$ we find is quite different from what would have
been predicted from the diffusion equation,
\beq
	\left( \partial_t - D_z\partial_z^2\right) n(z-vt) = 0,
\eeq
where $D_z$ is the usual position-space diffusion coefficient of order $1/T$,
not to be confused with our $D$.  The solution has the form
\beq
	n(z-vt) = n_0 \exp(-v(z-vt)/D_z)
\label{diffn}
\eeq
for $z>vt$, the region in front of the wall.  Thus in the small $v$ limit
the distribution approaches a constant in space, whereas ours has a finite
extent (albeit vanishing height) even as $v\to 0$.  This leads to at least one
extra power of $v^{-1}$ for the baryon asymmetry in the diffusion equation
approach, due to the integration over $z$ of the density in
(\ref{diffn}).\footnote{In fact there should be a second factor of $v^{-1}$
because by Fick's law, the flux is $J = -D_z dn/dz = vn_0$ at the bubble wall,
so $n_0$ is the flux at the wall divided by $v$.  This extra factor does not
appear in refs.\ \cite{JPT} or \cite{Heckler}.}

How do we understand the discrepancy between the two approaches?  To bolster
our confidence we can look for the same behavior in a simplified version of
the same physical system, by working in one dimension and taking the infinite
temperature limit, holding $D$ fixed.  Then eq.~(\ref{FP2}) becomes
\beq
        \left(\partial_p^2
        \mp {kp\over D|p|} \right)g_{k}^\pm =  0,
\label{FP3}
\eeq
whose solutions are mere exponential and trignometric functions.  In this case
the integral for $A^+_k$ analogous to (\ref{Akn}) can be done explicitly and
$\delta n(z)$ can be reduced to a single integral over $u\equiv (k/D)^{1/2}$,
\beq
	\delta n(z)  = {2N\over \pi}{d\over dw}\int_0^\infty {du\over u}
	e^{-zu^2}
	\left({w+u\over w^2+u^2}-{1\over w+u}\right),
\label{dn2}
\eeq
whose large distance behavior goes like $z^{-1/2}$.  Again we see that
the profile is not constant in position space even at zero velocity (although
the length scale $T^2/D$ we had previously is now gone because of taking the
$T\to\infty$ limit).  However, we find that if we replaced the boundary
condition (\ref{bc1}) at the wall by one that did not depend on momentum, by
letting the source term $S(p_z)$ be a constant (let $w\to 0$), then the
solution for $A_k$ would be a delta function at $k=0$, and eq.\ (\ref{dn2})
would indeed give a constant in $z$-space!  So we have discovered one reason
the two methods disagree: the Boltzmann equation demands that the boundary
condition for the distributions at the bubble wall are functions of momentum,
whereas the diffusion equation has integrated over the momenta from the outset,
and so cannot account for such details of the distribution.  Of course the full
Boltzmann equation should be the more accurate of the two methods.  It would be
interesting to compare the two predictions with the Monte Carlo method of CKN,
but since the latter display data only for a single small value of $v$, it is
impossible to measure their $v$-dependence in the region of $v=0.1$.

In summary, we have given an analytic expression for the baryon asymmetry due
to quarks reflecting from domain walls during the first order electroweak phase
transition, by solving the Boltzmann equation for the reflected particles.  The
solution is valid for slowly moving walls and temperatures which are large
compared to the quark masses or the width in momentum space for the reflection
asymmetry between the quarks and antiquarks.  Although the top quark mass is
actually not small compared to the critical temperature, the fact that we
obtained quantitative agreement with the previous results of CKN for $m/T$ as
large as unity is encouraging since we don't expect it to be greater than this.
Furthermore $m$ vanishes (up to thermal loop contributions) in the symmetric
phase outside the bubble, the region most important to the estimate of baryon
production, so it might be hoped that the small-mass approximation is better
than expected parametrically.  An interesting feature of our expression is its
insensitivity to the velocity of the expanding bubble wall, relative to
predictions based on the diffusion equation.

Our result makes it possible to estimate the baryon asymmetry $\Delta B$
directly in terms of the quark reflection asymmetry $\Delta{\cal R}(p_z)$, an
independent calculation which is being repeated \cite{CKV} because of
apparently conflicting calculations in the literature \cite{CKN}\cite{Fun}.
The quantitative results of electroweak baryogenesis through quark reflection
hinge upon this since the momentum-space width of $\Delta{\cal R}(p_z)$, on
which $\Delta B$ depends linearly, falls quite sharply with increasing quark
mass, whereas for small masses $\Delta B$ is suppressed by a factor of $m^3$.
Further application of the present work will be given in \cite{CKV}.

I would like to thank Larry McLerran and Sonia Paban for important
contributions during the early stage of this work, which was supported in part
by DOE grant DE-AC02-83ER-40105.  I also thank Kimmo Kainulainen and Alejandro
Ayala for helpful discussions.

\newpage
\unitlength=1.00mm
\begin{picture}(100.00,150.00)
\put(30.00,60.00){\line(1,0){60.00}}
\put(30.00,60.00){\line(0,1){80.00}}
\put(30.00,140.00){\makebox(0,0)[cc]{$-$}}
\put(30.00,130.00){\makebox(0,0)[cc]{$-$}}
\put(30.00,120.00){\makebox(0,0)[cc]{$-$}}
\put(30.00,110.00){\makebox(0,0)[cc]{$-$}}
\put(30.00,100.00){\makebox(0,0)[cc]{$-$}}
\put(30.00,90.00){\makebox(0,0)[cc]{$-$}}
\put(30.00,80.00){\makebox(0,0)[cc]{$-$}}
\put(30.00,70.00){\makebox(0,0)[cc]{$-$}}
\put(40.00,60.00){\makebox(0,0)[cc]{$|$}}
\put(50.00,60.00){\makebox(0,0)[cc]{$|$}}
\put(60.00,60.00){\makebox(0,0)[cc]{$|$}}
\put(70.00,60.00){\makebox(0,0)[cc]{$|$}}
\put(80.00,60.00){\makebox(0,0)[cc]{$|$}}
\put(90.00,60.00){\makebox(0,0)[cc]{$|$}}
\put(100.00,60.00){\makebox(0,0)[cc]{$z_0 k_0$}}
\put(20.00,140.00){\makebox(0,0)[cc]{$\phantom{-}0$}}
\put(20.00,130.00){\makebox(0,0)[cc]{$-1$}}
\put(20.00,120.00){\makebox(0,0)[cc]{$-2$}}
\put(20.00,110.00){\makebox(0,0)[cc]{$-3$}}
\put(20.00,100.00){\makebox(0,0)[cc]{$-4$}}
\put(20.00,90.00){\makebox(0,0)[cc]{$-5$}}
\put(20.00,80.00){\makebox(0,0)[cc]{$-6$}}
\put(20.00,70.00){\makebox(0,0)[cc]{$-7$}}
\put(20.00,60.00){\makebox(0,0)[cc]{$-8$}}
\put(40.00,50.00){\makebox(0,0)[cc]{$1$}}
\put(50.00,50.00){\makebox(0,0)[cc]{$2$}}
\put(60.00,50.00){\makebox(0,0)[cc]{$3$}}
\put(70.00,50.00){\makebox(0,0)[cc]{$4$}}
\put(80.00,50.00){\makebox(0,0)[cc]{$5$}}
\put(90.00,50.00){\makebox(0,0)[cc]{$6$}}
\put(30.00,140.00){\makebox(0,0)[cc]{$\bullet$}}
\put(40.00,118.67){\makebox(0,0)[cc]{$\bullet$}}
\put(50.00,101.67){\makebox(0,0)[cc]{$\bullet$}}
\put(60.00,86.33){\makebox(0,0)[cc]{$\bullet$}}
\put(70.00,72.33){\makebox(0,0)[cc]{$\bullet$}}
\put(80.00,58.67){\makebox(0,0)[cc]{$\bullet$}}
\bezier{384}(80.00,58.67)(35.67,120.33)(30.00,140.00)
\put(30.00,150.00){\makebox(0,0)[cc]{$\ln(s)$}}
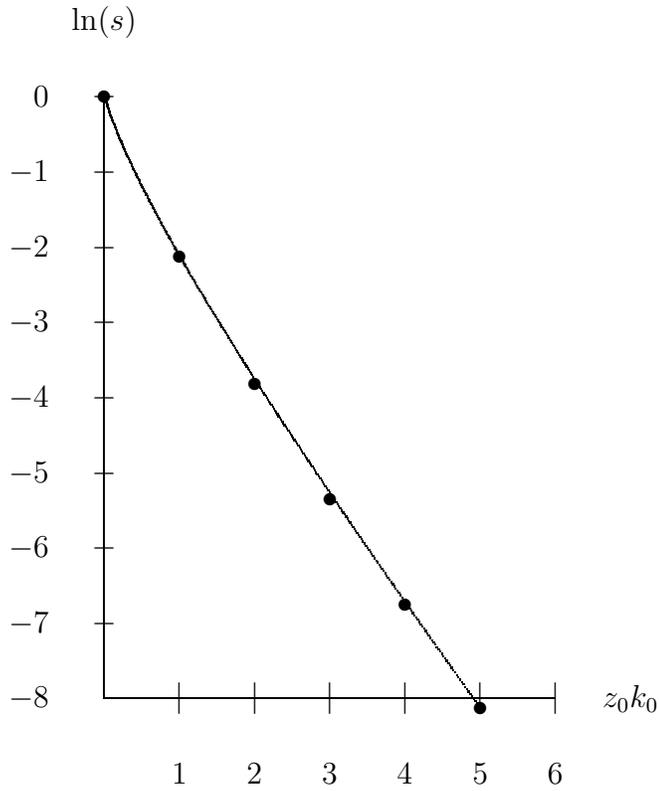
\put(30.00,40.00){\makebox(0,0)[lt]{\vbox{\hsize=3in Fig.~1: logarithm of the
suppression factor due to finite wall thickness. $z_0$ is proportional to the
bubble wall thickness (see text), and is in units of the diffusion length
$k_0^{-1}\cong 25 T^{-1}$.}}}
\end{picture}

\newpage
\thicklines
\begin{picture}(150.00,160.00)
\put(30.00,150.00){\makebox(0,0)[cc]{$\bullet$}}
\put(50.00,49.33){\makebox(0,0)[cc]{$\bullet$}}
\put(70.00,28.00){\makebox(0,0)[cc]{$\bullet$}}
\put(90.00,17.00){\makebox(0,0)[cc]{$\bullet$}}
\put(110.00,11.00){\makebox(0,0)[cc]{$\bullet$}}
\put(130.00,7.33){\makebox(0,0)[cc]{$\bullet$}}
\put(29.66,150.00){\line(0,-1){150.00}}
\put(29.66,0.00){\line(1,0){110.67}}
\put(40.00,69.00){\makebox(0,0)[cc]{$\bullet$}}
\put(60.00,36.00){\makebox(0,0)[cc]{$\bullet$}}
\put(80.00,21.67){\makebox(0,0)[cc]{$\bullet$}}
\put(100.00,13.67){\makebox(0,0)[cc]{$\bullet$}}
\put(119.67,9.00){\makebox(0,0)[cc]{$\bullet$}}
\put(35.00,87.67){\makebox(0,0)[cc]{$\bullet$}}
\bezier{424}(30.00,150.00)(30.00,79.33)(49.60,49.60)
\put(140.00,6.00){\makebox(0,0)[cc]{$\bullet$}}
\bezier{452}(140.00,6.00)(69.33,13.00)(49.60,49.60)
\put(29.67,150.00){\makebox(0,0)[cc]{$-$}}
\put(30.00,140.00){\makebox(0,0)[cc]{$-$}}
\put(30.00,130.00){\makebox(0,0)[cc]{$-$}}
\put(30.00,120.00){\makebox(0,0)[cc]{$-$}}
\put(30.00,110.00){\makebox(0,0)[cc]{$-$}}
\put(30.00,100.00){\makebox(0,0)[cc]{$-$}}
\put(30.00,80.00){\makebox(0,0)[cc]{$-$}}
\put(30.00,90.00){\makebox(0,0)[cc]{$-$}}
\put(30.00,70.00){\makebox(0,0)[cc]{$-$}}
\put(30.00,60.00){\makebox(0,0)[cc]{$-$}}
\put(30.00,50.00){\makebox(0,0)[cc]{$-$}}
\put(30.00,40.00){\makebox(0,0)[cc]{$-$}}
\put(30.00,30.00){\makebox(0,0)[cc]{$-$}}
\put(30.00,20.00){\makebox(0,0)[cc]{$-$}}
\put(30.00,10.00){\makebox(0,0)[cc]{$-$}}
\put(40.00,0.00){\makebox(0,0)[cc]{$|$}}
\put(50.00,0.00){\makebox(0,0)[cc]{$|$}}
\put(60.00,0.00){\makebox(0,0)[cc]{$|$}}
\put(70.00,0.00){\makebox(0,0)[cc]{$|$}}
\put(80.00,0.00){\makebox(0,0)[cc]{$|$}}
\put(90.00,0.00){\makebox(0,0)[cc]{$|$}}
\put(100.00,0.00){\makebox(0,0)[cc]{$|$}}
\put(110.00,0.00){\makebox(0,0)[cc]{$|$}}
\put(120.00,0.00){\makebox(0,0)[cc]{$|$}}
\put(130.00,0.00){\makebox(0,0)[cc]{$|$}}
\put(140.00,0.00){\makebox(0,0)[cc]{$|$}}
\put(20.00,150.00){\makebox(0,0)[cc]{$15$}}
\put(20.00,100.00){\makebox(0,0)[cc]{$10$}}
\put(20.00,50.00){\makebox(0,0)[cc]{$5$}}
\put(20.00,-10.00){\makebox(0,0)[cc]{$0$}}
\put(80.00,-10.00){\makebox(0,0)[cc]{$0.5$}}
\put(140.00,-10.00){\makebox(0,0)[cc]{$1.0$}}
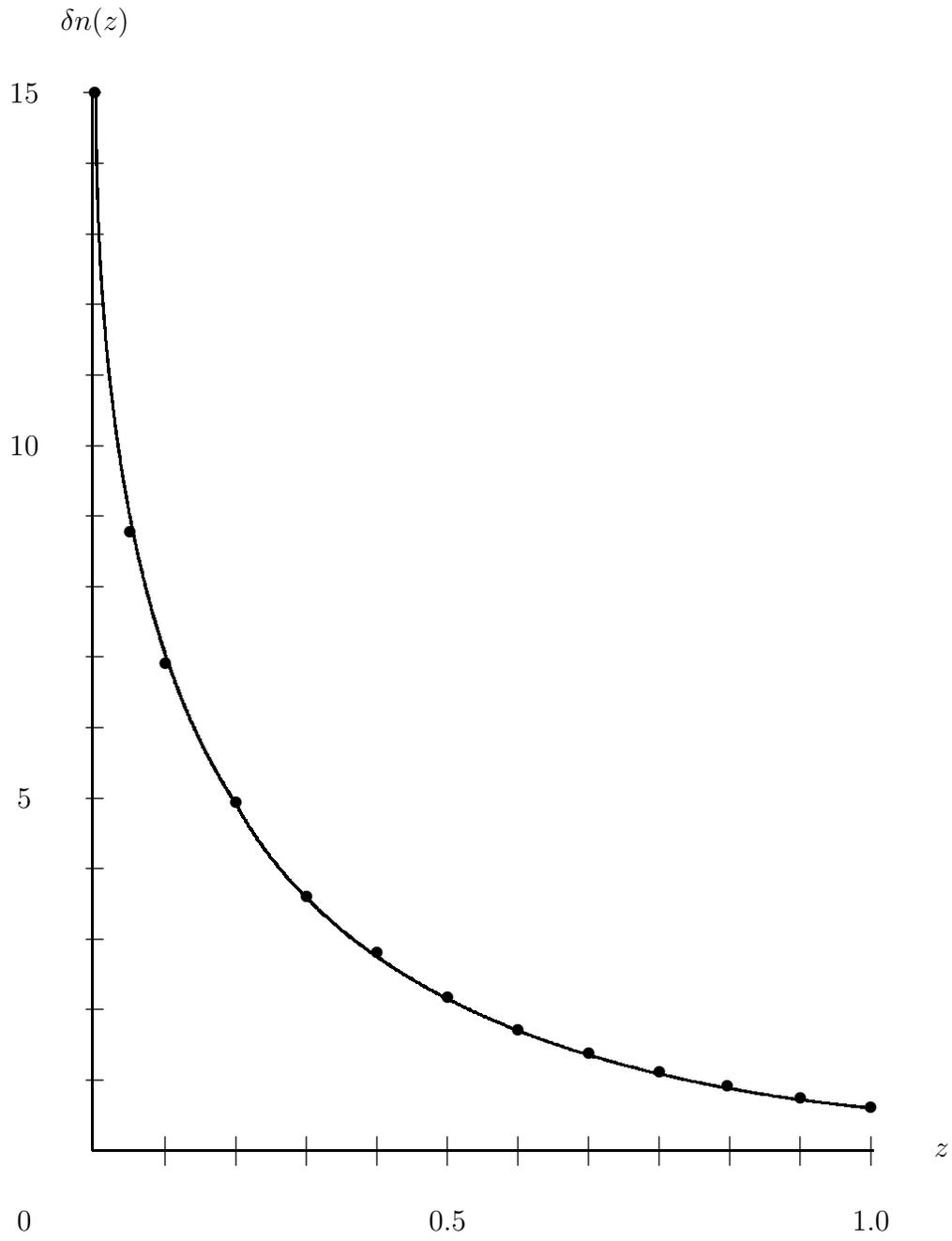
\put(30.00,-20.00){\makebox(0,0)[lt]{\vbox{\hsize=4.5in Fig.~2: The reflected
quark asymmetry profile in arbitrary units, versus distance in front of the
wall, in units of $k_0^{-1}\cong 25 T^{-1}$.}}}
\put(30.00,160.00){\makebox(0,0)[cc]{$\delta n(z)$}}
\put(150.00,0.00){\makebox(0,0)[cc]{$z$}}
\end{picture}

\end{document}